\documentclass[9pt,twocolumn,twoside]{extarticle}

\usepackage{authblk}

\usepackage{inputenc}
\usepackage[twoside,%
            letterpaper,includeheadfoot,%
            layoutsize={8.125in,10.875in},%
            layouthoffset=0.1875in,%
            layoutvoffset=0.0625in,%
            left=38.5pt,%
            right=43pt,%
            top=43pt,%
            bottom=32pt,%
            headheight=0pt,%
            headsep=10pt,%
            footskip=25pt,
            marginparwidth=38pt]{geometry}

\usepackage{newtxtext,newtxmath}

\usepackage[english]{babel}
\usepackage{ae,aecompl}

\usepackage{algorithm}

\usepackage{graphicx}
\usepackage{mwe}
\usepackage[colorlinks=true, allcolors=blue]{hyperref}

\usepackage{commath}
\usepackage{amsmath}

\usepackage{soul}

\usepackage{amssymb}
\usepackage{mathrsfs}
\usepackage{mathtools}
\usepackage{booktabs}
\usepackage{siunitx}
\usepackage{comment}
\usepackage{savesym}
\usepackage{siunitx}
\usepackage{pythonhighlight}
\usepackage{physics}

\usepackage[numbers,sort&compress,merge,round]{natbib}
\usepackage{jabbrv}
\usepackage{etoolbox}

\usepackage{tikz}
\savesymbol{plate}
\usetikzlibrary{bayesnet}

\usepackage{xcolor}

\newcommand\vx{\vb x}
\newcommand\vy{\vb y}
\newcommand\vz{\vb z}
\newcommand{\varo}{\vb*{\theta}_1}
\newcommand{\vart}{\vb*{\theta}_2}

\usepackage{nameref}%
 \GetTitleStringSetup{expand}
\usepackage{cleveref}

\title{{\huge\textbf{A Bayesian neural network predicts the dissolution of compact planetary systems}}}

\author[a,$\dagger$]{Miles Cranmer}
\author[a]{Daniel Tamayo}
\author[b,c]{Hanno Rein}
\author[d]{Peter Battaglia}
\author[e]{Samuel Hadden}
\author[f,g]{\\Philip J. Armitage}
\author[g,a,h]{Shirley Ho}
\author[g,a]{David N. Spergel}

\affil[a]{Princeton University, Princeton, New Jersey, USA}
\affil[b]{University of Toronto at Scarborough, Toronto, Ontario, Canada}
\affil[c]{University of Toronto, Toronto, Ontario, Canada}
\affil[d]{DeepMind, London, UK}
\affil[e]{Center for Astrophysics | Harvard \& Smithsonian, Cambridge, Massachusetts, USA} 
\affil[f]{Stony Brook University, Stony Brook, New York, USA}
\affil[g]{Flatiron Institute, New York, New York, USA}
\affil[h]{Carnegie Mellon University, Pittsburgh, Pennsylvania, USA}
\affil[$\dagger$]{{\normalfont Corresponding author. Email: mcranmer@princeton.edu}}

\date{\vspace{-2em}}

\renewcommand\v{\mathbf}

\renewenvironment{abstract}
 {\small
  \begin{center}
  \bfseries \abstractname\vspace{-.5em}\vspace{0pt}
  \end{center}
  \list{}{%
    \setlength{\leftmargin}{20mm}%
    \setlength{\rightmargin}{\leftmargin}%
  }%
  \item\relax}
 {\endlist}

\begin{document}

\newcommand\materials{the appendix\xspace}

\twocolumn[
\begin{@twocolumnfalse}
\maketitle
\begin{abstract}
\normalsize
Despite over three hundred years of effort, no solutions exist for predicting when a general planetary configuration will become unstable. We introduce a deep learning architecture to push forward this problem for compact systems. While current machine learning algorithms in this area rely on scientist-derived instability metrics, our new technique learns its own metrics from scratch, enabled by a novel internal structure inspired from dynamics theory. Our Bayesian neural network model can accurately predict not only if, but also when a compact planetary system with three or more planets will go unstable. Our model, trained directly from short N-body time series of raw orbital elements, is more than two orders of magnitude more accurate at predicting instability times than analytical estimators, while also reducing the bias of existing machine learning algorithms by nearly a factor of three. Despite being trained on compact resonant and near-resonant three-planet configurations, the model demonstrates robust generalization to both non-resonant and higher multiplicity configurations, in the latter case outperforming models fit to that specific set of integrations. The model computes instability estimates up to five orders of magnitude faster than a numerical integrator, and unlike previous efforts provides confidence intervals on its predictions. Our inference model is publicly available in the \texttt{SPOCK}$^\ast$ package, with training code open-sourced$^\ddag.$
\vspace{1em}
\end{abstract}
\end{@twocolumnfalse}
]

\section{Introduction}
The final growth of terrestrial bodies in current theories of planet formation occurs in a phase of giant impacts \citep{kokubo12}\footnotetext{$\ast$ \url{https://github.com/dtamayo/spock}}\footnotetext{$\ddag$ \url{https://github.com/MilesCranmer/bnn_chaos_model}}.\setcounter{footnote}{3}
During this stage, the number of planets slowly declines as bodies collide and merge \citep{volk15, pu15}.
Close planetary encounters and the wide dynamic range exhibited by the times between consecutive collisions computationally limit current numerical efforts to model this process.
Two theoretical roadblocks impede the development of a more efficient iterative map for modeling planet formation.
First, one must predict a distribution of instability times from a given initial orbital configuration.
Second, one must predict a distribution of post-collision orbital architectures \citep[e.g.,][]{tremaine15} subject to mass, energy and angular momentum constraints.
Towards this end, we focus on the long-standing former question of instability time prediction.

In the compact dynamical configurations that characterize the planet formation process, the simpler two-planet case is understood analytically.
In this limit, instabilities are driven by the interactions between nearby mean-motion resonances (MMRs), i.e., integer commensurabilities between the orbital periods of the planets like the 3:2 MMR between Pluto and Neptune \citep{wisdom80, deck13, petit18, hadden18}.
While the general higher-multiplicity case is not yet understood, two important results guide our analysis and provide an important test for any model.
First, when planets are initialized on circular orbits, chaos is driven by the overlap of 3-body MMRs between trios of adjacent planets \citep{quillen11}, and theoretical estimates of the timescale required for the orbits to reach orbit-crossing configurations accurately match numerical integrations \citep{petit20}.
As we show below, such analytical estimates perform poorly in the generic eccentric case where the effects of 2-body MMRs are dominant \citep{petit20, tamayo20}.
However, analytical and empirical studies agree that while the dynamical behavior changes strongly from the two to three-planet case \citep{chambers96, yoshinaga99, marzari02, zhou07, faber07, smith09, matsumoto12, pu15}, three-planet systems are the simplest prototype for predictions at higher multiplicities in compact systems \citep{tamayo20, petit20}. 

We recently presented a machine learning model, dubbed the Stability of Planetary Orbital Configurations Klassifier, or SPOCK, trained to classify the stability of compact planetary systems over timescales of $10^9$ orbits \citep{tamayo20}.
This represented a long-term effort to exploit the substantial but incomplete current analytical understanding  \citep{chirikov79, wisdom80, deck13, hadden18, hadden19} to engineer summary metrics that captured these systems' chaotic dynamics; these features were then used by the machine learning model to classify whether the input configuration would be stable over $10^9$ orbits.

While simple binary stability classification is effective for constraining physical and orbital parameters consistent with long-term stability \citep{tamayo20b}, other applications like modeling terrestrial planet formation require the prediction of continuous instability times.
Additionally, several fields in which it is challenging to find effective hand-picked features---such as computer vision, speech recognition, and text translation---have been revolutionized by neural networks in the last decade \citep[notable early breakthroughs include][]{alexnet,speech,seq2seq}.
Rather than relying on domain expert input, these flexible models learn data-driven features that can often significantly outperform human-engineered approaches.
A key theme with deep learning models is that their structure resembles the hand-designed algorithm, but with added flexibility parametrized by neural networks \citep[for discussion, see][]{deeplearn}.
For example, modern computer vision models consist of learned convolutional filters which take the place of hand-designed filters in classic algorithms \cite{sift}.

Pursuing a deep learning approach, we present a neural network that, trained only on short-time series of the orbits in compact planetary systems, not only improves on long-term predictions of previous models based on engineered features \citep{tamayo16, tamayo20}, but also significantly reduces the model bias and improves generalization beyond the training set. 
We design our model as a Bayesian neural network, which
naturally incorporates confidence intervals into its
instability time predictions, accounting both for model uncertainty as well
as the intrinsic uncertainty due to the chaotic dynamics.
Finally, unlike previous machine learning models based on decision trees
\citep{tamayo16, tamayo20}, our model is differentiable. 
That is, we can extract from the model estimates of the derivatives of the
predicted instability times with respect to the parameters defining the orbital configuration in question.
Such gradient information can significantly speed up parameter
estimation using Hamiltonian Monte Carlo techniques \citep{agol20}.

\section{Model}
\label{sec:model}
\subsection{Dataset generation}
\label{sec:datagen}

We focus on the regime leading to typical compact multi-planet systems observed to date, with mass ratios with the central star ranging from $10^{-7}$ (roughly the ratio of the Moon-mass embryos thought to initially characterize the giant impact phase, relative to the Sun) to $10^{-4}$ (roughly Neptune's mass relative to the Sun).
As detailed in \materials, we place planets
on nearly coplanar orbits, with adjacent planets spaced within 30 mutual Hill radii\footnote{The mutual Hill radius $R_H$ is a relevant length scale within which the gravity of the planets dominates that of the star $R_H \approx a_1 (\mu_1 + \mu_2)^{1/3}$, where $a_1$ is the inner planet's semi-major axis, and $\mu_1$ and $\mu_2$ are each planet's mass ratio relative to the star.} of one another \citep[e.g.,][]{weiss18}.
Orbital eccentricities in observed systems are often poorly constrained,
so we consider the range from initially circular to orbit-crossing values.

We train our model on the set of 113,543 publicly available,
compact three-planet configurations in and near strong resonances from \cite{tamayo20}.
In particular, this ``resonant'' dataset initializes one pair of planets in or near a strong MMR, while the third planet's orbital parameters are chosen randomly.
This choice focuses the training set on the narrow resonant regions of phase space where the dynamical behavior changes most strongly, and we later test the model's generalization to non-resonant systems in \cref{sec:results}.

Each initial condition was integrated for $10^9$ orbits of the innermost planet using the
\texttt{WHFast} integrator \citep{reintamayo15} in the REBOUND N-body package \citep{rein12}. 
If at any point two planets came within a distance of one another given by the sum of their Hill radii,
the simulation was stopped
and the instability time was recorded.
Because gravity is scale invariant, the instability time $t_\text{inst}$ is most usefully non-dimensionalized by the innermost orbital period $P_\text{orb}$.
Given the large dynamic range in timescales over which instabilities can occur, we define the dimensionless log instability time $T \equiv \log_{10}(t_\text{inst}/P_\text{orb})$.
Configurations with instability times longer than
$10^9$ orbits ($T>9$) were labeled as stable, and integration
was stopped.

\renewcommand{\paragraph}[1]{\noindent\textbf{#1.}}

\subsection{Network architecture}
\label{sec:architecture}

\begin{figure*}
    \centering
    \includegraphics[width=0.8\linewidth]{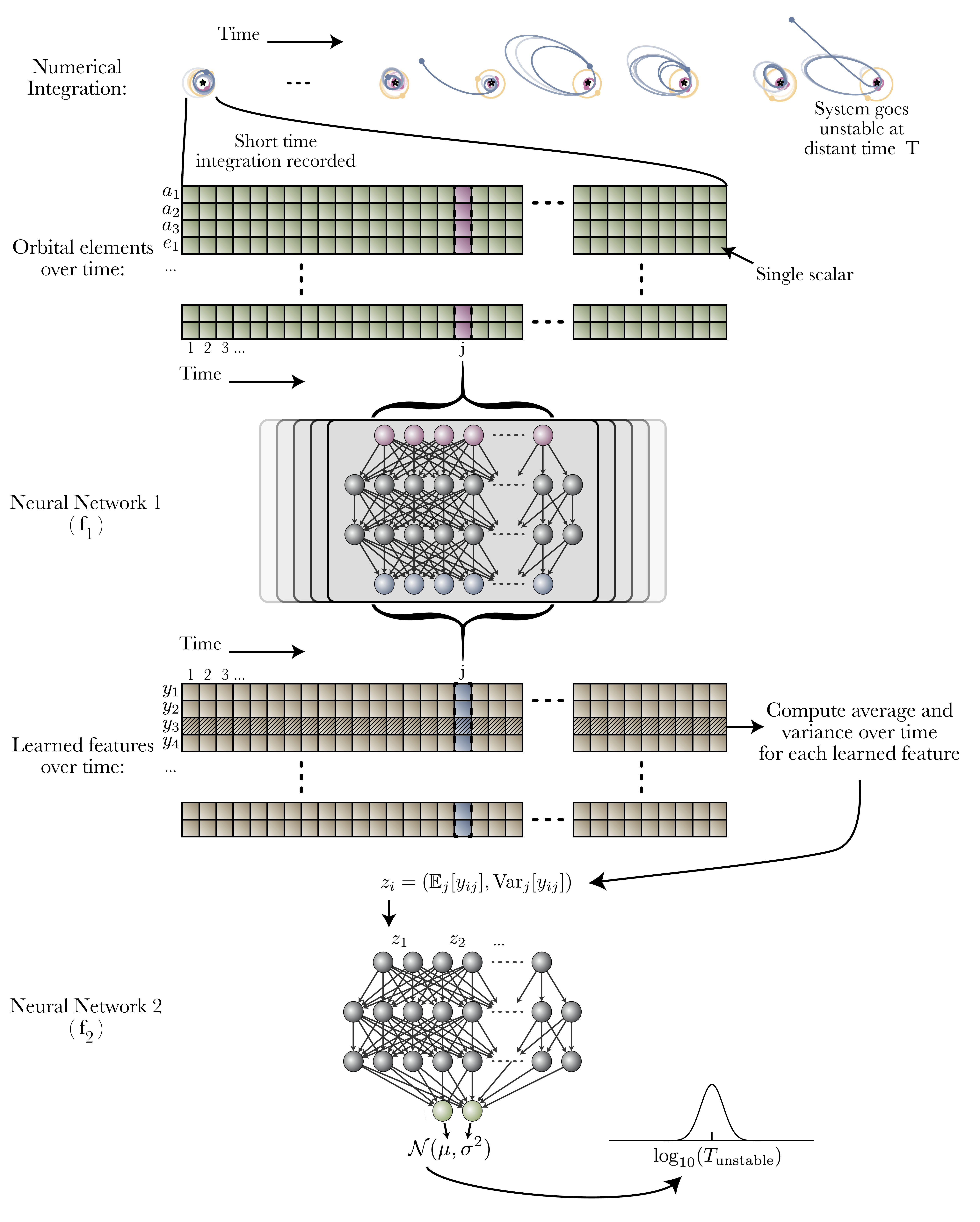}
    \caption{Schematic of our model. The model
    numerically integrates 10,000 orbits for a compact
    3-planet system (top) and records orbital elements at 100 times.
    Neural network $f_1$ creates learned summary features from these elements at each time.
    Neural network $f_2$ takes the
    average and variance of these features
    as input and estimates
    a distribution over possible instability times.
    }
    \label{fig:setup}
\end{figure*}

To predict systems' instability times, we perform a computationally cheap numerical integration of the first $10^4$ orbits, and use this time series to make long-term predictions.
Each of the three planets' 3D positions and velocities correspond to six standard orbital elements (\materials), which we record at $n_t=100$ equally spaced outputs across the short integration. 
In addition we pass the three constant mass ratios for each planet relative to the star, as well as the time value, for a combined input matrix of real values $X \in \mathbb{R}^{3 + 19 n_t}$ for a given configuration.

Because the dynamics of compact multi-planet systems are chaotic, instability times for a given initial orbital configuration are  effectively non-deterministic.
Nevertheless, numerical experiments \citep{rice18, hussain19}
have shown that instability times for unstable, compact multi-planetary systems settle to well defined, approximately log-normal distributions.
Thus, rather than predicting a single instability time for a given orbital configuration,
our model maps from an input initial orbital configuration to a predicted log-normal distribution of instability times, i.e., a Gaussian distribution of $T$ with mean $\mu$ and variance $\sigma^2$.
This gives the network the flexibility to both model the fundamental uncertainties imposed by the chaotic dynamics, and to incorporate model uncertainty into its predictions by assigning larger widths to configurations it is less sure about. 

In our initial efforts, we experimented with various combinations of convolutional neural networks \citep[see reviews by][]{cnn0,cnn1,cnnreview}, long short-term memory networks \cite{lstm}, 1D scattering transforms \citep{wavelet}, regular multi-layer perceptrons \citep[MLP, see][]{deeplearn} and gaussian processes \cite{gpbook}.
All of these models underperformed or tended to overfit the data.

The fundamental challenge for making such predictions is the sharp transitions in dynamical behavior at MMRs, where instability times can change by several orders of magnitude \citep{obertas17} over sub-percent changes in planetary orbital periods, i.e., in the original space of orbital elements.
We found substantially improved performance by structurally splitting the problem into three components:
1) Find a transformation from the sharply punctuated space of orbital elements to new variables.
2) Calculate statistical summaries of the time series in these transformed variables.
3) Use these summary features to predict a log-normal distribution over instability times for the input orbital configuration, parametrized by mean $\mu$ and variance $\sigma^2$. This is illustrated in \cref{fig:setup}.

We model steps 1 and 3 with neural networks $f_1$ and $f_2$, respectively.
In step 2, we choose to use a well-motivated but non-standard pooling operation, and calculate both the mean and variance for each transformed time series (deep neural networks traditionally use one of either sum, mean, or max pooling).
This structure was motivated by the hand-engineered approach of \cite{tamayo20}, whose features are means and variances of theoretically-motivated instability metrics. 
In this case, we give the machine learning model additional flexibility to learn its own instability metrics.
This design process is analogous to how the invention of convolutional neural networks was motivated by creating similar structures to hand-engineered algorithms which convolve filters over an image. 

\subsection{Likelihood}
\label{sec:like}

Our model is parametrized by $m$ neural network weights $\theta\equiv (\varo, \vart), \theta\in\mathbb{R}^m$ ($\varo$ for $f_1$
and $\vart$ for $f_2$).
Defining the training set $D$ as the collection of input orbital configurations and their associated N-body instability times, we seek the most likely set of model parameters given the data, i.e., we maximize $P(\theta|D)$, which is in turn proportional to $P(D|\theta)P(\theta)$.

Our model predicts a log-normal distribution of instability times for any input orbital configuration.
For a given set of network weights $\theta$, the likelihood $P(D|\theta)$ is then simply the product of the probabilities that each training set example's output $T_i$ is drawn from the associated Gaussian $\mathcal{N}(\mu_i, \sigma_i^2)$ predicted by the model. 
As discussed above, this choice is motivated by the numerical result that the distribution in $T$ is normal, for different configurations with a wide range of mean instability times \citep{hussain19}.

Note that we have $4 < T \leq 9$ as a constraint for unstable simulations: $T<4$ simulations
are not included in the training set,
and $T>9$ integrations were terminated at $T=9$ and have an unknown $T$. 
Thus, we build a truncated normal
distribution with a cutoff at $T=4$, and the cumulative
probability of the Gaussian above $T=9$ being counted
towards a classification of stability.
A mathematical derivation of this likelihood is given in
\materials.

$P(\theta)$ is a prior on the neural network's parameters.
The per-parameter prior here is unimportant:
what matters is the prior
\textit{induced on the output of the model}, and we
use an uninformative Gaussian prior on parameters to
induce an uninformative prior on the output.
See \cite{wilsoncase} for a detailed discussion of priors in Bayesian
deep learning.

\subsection{Bayesian neural network implementation}
\label{sec:training}

By having our model predict a distribution of instability times with a finite width, we account for intrinsic uncertainty (sometimes referred to as ``aleatoric'' uncertainty).
However, we also wish to include extrapolation uncertainty (or ``epistemic'' uncertainty) for systems that differ from those found in the training set.
To do this we marginalize over potential model parameters, with what is referred to as a ``Bayesian Neural Network'' (BNN).
This is a neural network whose parameters are distributions rather than point values, and is trained with Bayesian inference.
These models naturally fold in extrapolation uncertainty via marginalization.

This concept is familiar in traditional statistical inference,
where one can marginalize out the internal nuisance parameters
of a model using Markov Chain Monte Carlo (MCMC) techniques.
The fact that neural networks typically have
millions of parameters renders MCMC computationally prohibitive,
and various practical simplifications are adopted for implementing a BNN.
The most common strategy is Monte Carlo dropout \cite{mcdropout,mcdropout2}
which treats the neural network's
parameters as independent Bernoulli random variables,
and has been used in several astronomical applications
\cite{egmcdropout1,egmcdropout2,egmcdropout3,egmcdropout4}.
A selection of other techniques includes
Bayes by Backprop \cite{bbb},
Bayesian Layers \cite{bayesianlayers},
variants of normalizing flows \citep[e.g., ][]{louizos},
Bayes by Hypernet \cite{bbh1,bbh2}, and many other strategies. 
One recently-proposed strategy, named ``MultiSWAG'' \cite{swag,multiswag} learns a distribution over the posterior of weights that best fit the training set, without a diagonal covariance assumption, and
is much closer to standard MCMC inference.
We experimented with a selection of common techniques:
Monte Carlo dropout, Bayes by Backprop, and MultiSWAG,
and found that ``MultiSWAG''
produced the best accuracy and uncertainty estimations 
on the validation dataset.

To move beyond a single best-fit set of parameters $\theta$, SWAG, or ``Stochastic Weight Averaging Gaussian'' \cite{swa,swag}, instead
fits a Gaussian to a mode of the posterior over $\theta$,
assuming a low-rank covariance matrix.
This was extended in \cite{multiswag} to ``MultiSWAG,''
which repeats this process for several
modes of the weight posterior, to help fill out the highly degenerate parameter space.
This technique is summarized below:

\begin{enumerate}
    \item Train $f_1$ and $f_2$ simultaneously via stochastic gradient descent until the parameters settle into a minimum of the weight posterior.
    \item Increase the learning rate and continue training. This causes the optimizer to take a random walk in parameter space near the minima, which is assumed to look like a high-dimensional Gaussian.
    \item Accumulate the average parameters along this random walk as well as a low-rank approximation of the covariance matrix.
    \item The average parameters not only provide better generalization performance (stochastic weight averaging or SWA), but we have additionally fit a Gaussian to a mode of the parameter posterior. We can thus sample weights from this Gaussian to marginalize over parameters. This is SWAG \citep{swag}.
    \item The next step is to repeat this process from a different random initialization of the parameters. This will find another mode of the parameter posterior.
    \item Fit $\sim$30 different modes of the parameter space. We can then sample weights from multiple modes of the parameter posterior, which gives us a more rigorous uncertainty estimates. This is MultiSWAG \cite{multiswag}.
\end{enumerate}

Training a neural network through stochastic gradient descent
is approximately the same as Bayesian sampling of the weights
\citep{sgdbayes2,sgdbayes}, so
this aforementioned process allows one to learn a Bayesian posterior
over the weights of a neural network $P_\text{MultiSWAG}(\theta)$.

Once we have learned $P_\text{MultiSWAG}$, we can draw from it to sample a set of network weights $\theta$. 
This model then predicts a log-normal distribution of instability times with mean $\mu$ and variance $\sigma^2$ for the given input orbital configuration, from which we can finally sample a log instability time T. 
We can write a forward model for this prediction as follows: 
\begin{align}
    (\varo, \vart) &\sim P_\text{MultiSWAG}(\theta), \label{eqn:modeleqn}
    \\
    \vb{y}_t &= f_{1}(\vb{x}_t; \varo) \qq{for all} \vb x_t \equiv X_{:\,, t},
    \label{eqn:modeleqn2}
    \\
    \vb{z} &\sim (\mathbb{E}_t[\vb{y}_t], \text{Var}_t[\vb{y}_t]),
    \label{eqn:modeleqn3}
    \\
    (\mu, \sigma^2) &= f_{2}(\vb{z}; \vart),
    \label{eqn:modeleqn4}
    \\
    T_\text{instability} &= 10^{T} \qq{for} T \sim \mathcal{N}(\mu, \sigma^2),
    \label{eqn:modeleqn5}
\end{align}
where $t$ is a time step from $1$ to $100$.
Here, we have labeled $\vb{y}_t$ as the learned transformed
variables for a single time step of the system (brown cells in \cref{fig:setup}), and $\vb z$
as the average and variance of these transformed variables over time.
To account for statistical errors due to our finite number of time series samples, we sample the $\mathbf{z}$ from normal distributions with frequentist estimates of the variance in the sample mean and variance: $\frac{\text{Var}_t[\mathbf{y}_t]}{n_t}$
and $\frac{2 \text{Var}_t[\mathbf{y}_t]^2}{(n_t - 1)}$, respectively.
A Bayesian graphical model for this is shown in \materials.
Repeatedly sampling in this way provides a predicted distribution of T given the input orbital configuration, marginalized over the posterior distribution of network weights $\theta$.

We split our data into 60/20/20\% train/validation/test,
train our model on 60\% of our $\approx 100,000$ training examples of resonant and near resonant systems,
and validate it on half of the remaining data to tune the hyperparameters.
Hyperparameters for our model are given in
the supplementary material, and we also release
the code to train and evaluate our model.

With this trained model, we then explore its performance
on the remaining 20\% holdout data from the resonant dataset, as well as other datasets described below.

\section{Results}
\label{sec:results}

\subsection{Resonant test dataset}
For a given orbital configuration, our probabilistic model produces one sample of $T$.
If a given sample is above $T=9$,
we treat the sample as a ``stable'' prediction. Since we are
unable to make specific time predictions above the maximum integration time in our training dataset of $T=9$,
we resample from a user-defined
prior $P(T|T\geq 9)$ for each occurrence.
For the purposes of this study, we assume a simple analytic form for this prior,
though followup work on this prior is ongoing (see supplementary).

For all results, we sample
10,000 predicted values of the posterior over
$T$ per planetary system.
We compare our predictions against several alternatives which
are explained below.
Since the models we compare against can only produce
point estimates while our model predicts a distribution,
we take the median of our model's predicted posterior over $T$.
This is used for plotting points, as well as for
computing root-mean-square prediction errors.

We first compute the N-body versus predicted (median) $T$ value over the holdout test dataset of $\approx 20,000$ examples not seen during training, which can be seen in the bottom middle panel of \cref{fig:reg}.
We reiterate that the N-body instability times measured for the various orbital configurations in our training set are not `true' instability times, but rather represent single draws from the different planetary systems' respective instability time distributions, established by their chaotic dynamics. 
To estimate a theoretical limit (bottom right panel of \cref{fig:reg}) we use the results from \cite{hussain19}, who find that the T values measured by N-body integrations (x-axis of \cref{fig:reg}) should be approximately normally distributed around the mean instability time predicted by an ideal model. 
We use a random standard deviation drawn from the values measured empirically for compact systems by \cite{hussain19}, which they find are sharply peaked around $\approx 0.43$ dex, independent of whether or not the system is near MMRs, and valid across a wide range of mean instability times. 
We plot this representative intrinsic width of 0.43 dex as dotted lines on all panels for comparison. 

While we defer a detailed comparison to previous work to the following section, we measure a root mean square error (RMSE) of 1.02 dex for our model on the holdout test set.
We note that while the RMSE is an intuitive
metric for comparing models, it does not provide a full picture
for a model that is trained on a different loss function to predict both $\mu$ and $\sigma^2$,
A model that can predict its own $\sigma^2$ will
sacrifice worse $\mu$ accuracy in challenging regions of parameter space
to better predict it on more easily predictable configurations.
For comparison, if we weight the RMSE by the predicted signal-to-noise ratio (SNR),  $\mu^2/\sigma^2$, the model achieves 0.87 dex, within a factor of $\approx 2$ of the theoretical limit. 
These uncertainties provide confidence estimates in the predicted values, and can indicate to a user when to invest in a computationally costly direct integration.
We apply transparency to our predictions in \cref{fig:reg} according to the model-predicted SNR, highlighting that the poorest predictions were typically deemed uncertain by the model.

Finally we quantitatively test whether the model-predicted uncertainties $\sigma$ accurately capture the spread of N-body times around the predicted mean values $\mu$.
For each test configuration, we predict $\mu$, subtract it from its respective $T$ measured by N-body integration, and divide by the predicted $\sigma$.
If this distribution approximates a Gaussian distribution
of zero mean and unit variance, the model's uncertainty estimates are accurate.
We find that a Komolgorov-Smirnov test cannot confidently distinguish our predictions from this ideal Gaussian (p-value of 0.056), and plot the two distributions in \cref{fig:exampledist} of the supplementary material. 

\subsection{Comparison to Previous Work}

Guided by the dynamical intuition that short-timescale instabilities are driven by the interaction of MMRs \citep{wisdom80, hadden18, tamayo20}, we chose to train our model on systems with particular period ratios and orbital elements in the narrow ranges near such resonances where the dynamical behavior changes sharply \citep{obertas17}.
It is therefore important to test how well such a model generalizes to a more uniform coverage of parameter space, given that most observed orbital architectures are not in MMRs (possibly because such configurations typically have short lifetimes and have been eliminated). 
Additionally, previous work has typically ignored the sharp variations near MMRs to fit overall trends in instability times \citep{obertas17}, so a test on resonant systems would not provide a fair comparison.

For this generalization test and comparison, we use the `random' dataset of \cite{tamayo20}, 25,000 three-planet systems with the same mass ratio and inclination distributions as above, and eccentricities drawn log-uniformly from $\approx 10^{-3}$ to orbit-crossing.
Rather than drawing near-integer period ratios as in our resonant training set, the spacing between adjacent planets is drawn uniformly between [3.5, 30] mutual Hill radii \citep[see][]{tamayo20}.

We find that our model exhibits only a minor loss in performance (1.20 vs 1.02 dex RMSE) generalizing to this uniform distribution of orbital configurations \cref{tbl:reg}. 
This supports the assertion that instabilities in compact systems within $10^9$ orbits are dominantly driven by the MMRs we focused on in our training sample \citep{tamayo20}.
To compare our results to the extensive set of past efforts, we divide previous approaches into three broad groups.

\newcommand\qar[1]{\begin{tabular}{c} #1 \end{tabular}}

\begin{table}
\begin{center}
    \begin{tabular}{@{}lcccc@{}}
    \toprule
    \textbf{Resonant}  &&&& \\
    \midrule
    Model & RMSE & \qar{Classif. \\ accur.} & \qar{Bias$^\dag$ \\ (4, 5)} & \qar{Bias \\ (8, 9)}\\
    \midrule
    Obertas et al. (2017) & 2.12 & 0.628 & 1.04 & -1.71 \\
    Petit et al. (2020)   & 3.22 & 0.530 & 3.99 & $\hphantom{\text{-}}$0.54 \\
    Tamayo et al. (2020) & 1.48 & 0.946 & 2.07 & -0.62 \\
    Modified$^\ast$ Tamayo+20 & 0.99 & 0.946 & 0.65 & -0.60 \\
    Ours                  & 1.02 & 0.952 & 0.29 & -0.38 \\
	Ours, SNR-weighted    & 0.87 & 0.971 & 0.18 & -0.25 \\
    \textit{Theoretical limit}  & \textit{0.43} &\textit{ 0.992} &\textit{ 0.05} &\textit{ -0.04 }\\
    \midrule
    \textbf{Random}  &&&& \\
    \midrule
    Obertas et al. (2017) & 2.41 & 0.721 & 2.15 & -0.93 \\
    Petit et al. (2020)   & 3.09 & 0.517 & 4.17 & $\hphantom{\text{-}}$0.50 \\
    Tamayo et al. (2020)                   & 1.24 & 0.949 & 1.16 & -0.59 \\
    Modified$^\ast$ Tamayo+20   & 1.14 & 0.945 & 0.79 & -0.70 \\
    Ours                  & 1.20 & 0.939 & 0.40 & -0.51 \\
	Ours, SNR-weighted    & 1.09 & 0.959 & 0.23 & -0.49 \\
    \textit{Theoretical limit    } &\textit{ 0.44} &\textit{ 0.989} &\textit{ 0.06} &\textit{ -0.04 }\\
    \midrule
     & (\si{dex}) & (AUC) & (\si{dex}) & (\si{dex})\\
    \midrule
        \multicolumn{5}{l}{$^\dag$Average difference between predicted minus true $T$ in given range}\\
        \multicolumn{5}{l}{$^\ast$Modified and retrained for regression.}\\
    \bottomrule
    \end{tabular}
\end{center}
    \caption{Statistical summaries of each estimator applied to a holdout test
    portion of the resonant dataset, and all of the random dataset.}
    \label{tbl:reg}
\end{table}

\newcommand\tmpwidth{.3\linewidth}
\begin{figure*}[!ht]
    \centering
  \begin{tabular}[b]{ccc}
    \includegraphics[width=\tmpwidth,trim={0 0 0 20pt},clip]{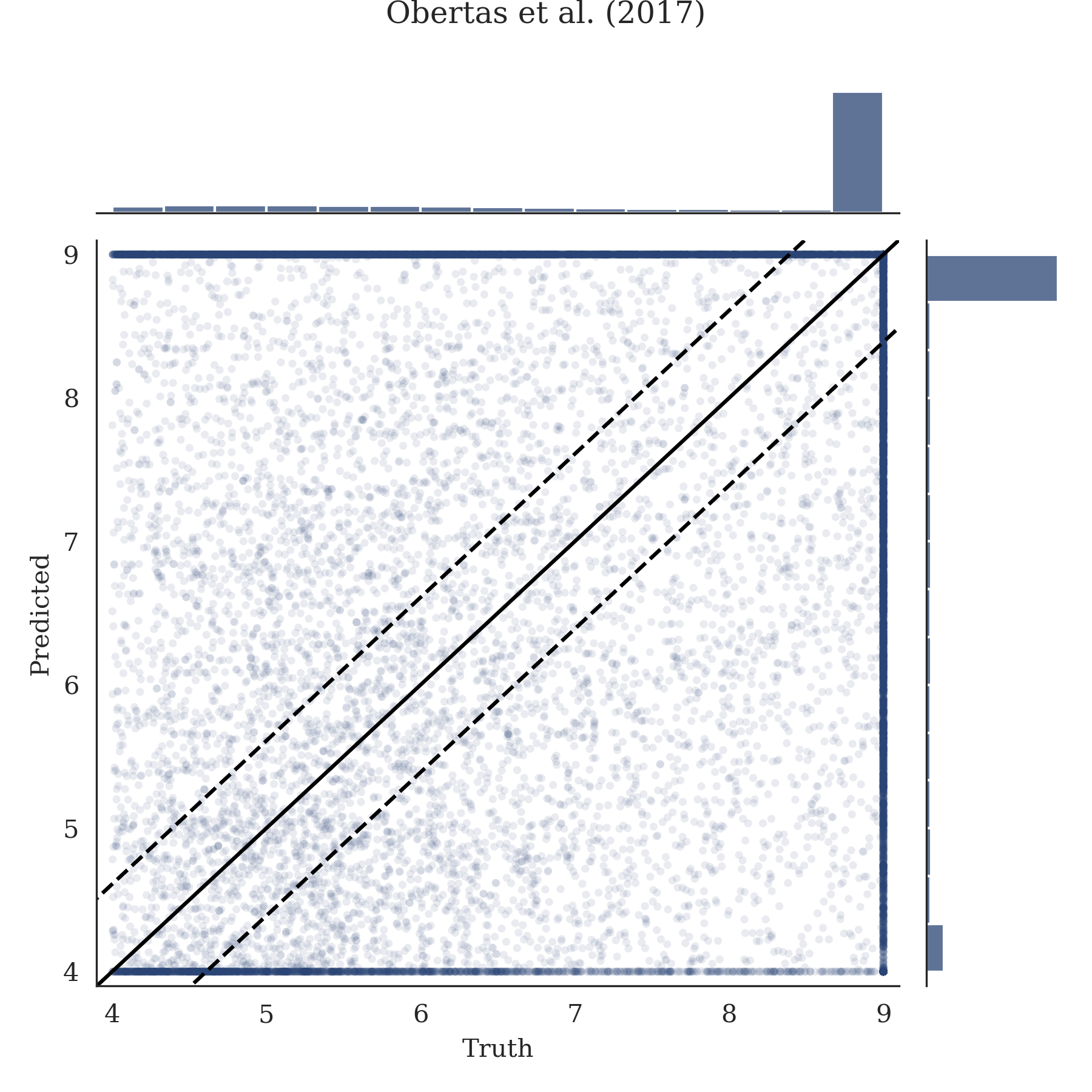} &
      \includegraphics[width=\tmpwidth,trim={0 0 0 20pt},clip]{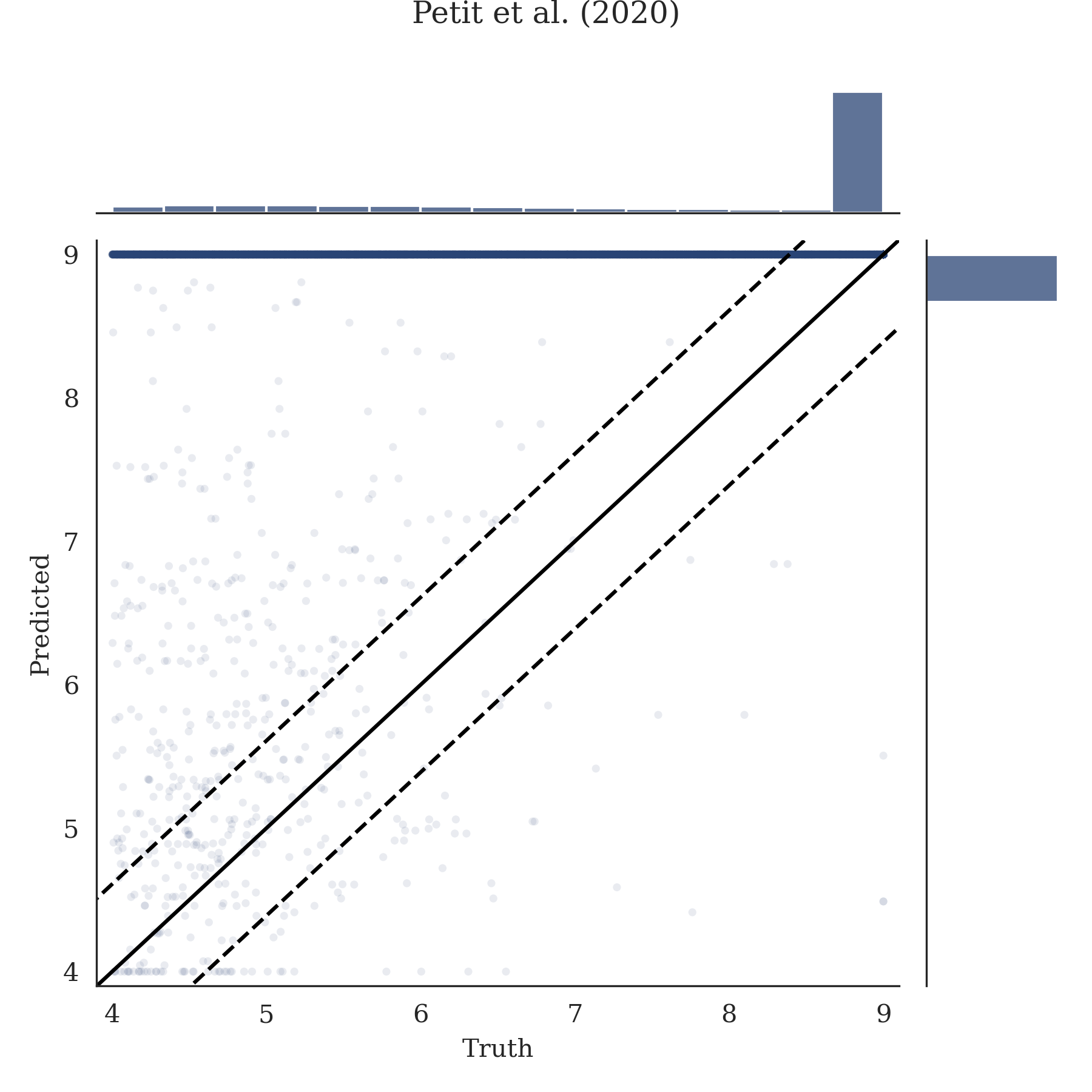} &
    \includegraphics[width=\tmpwidth,trim={0 0 0 20pt},clip]{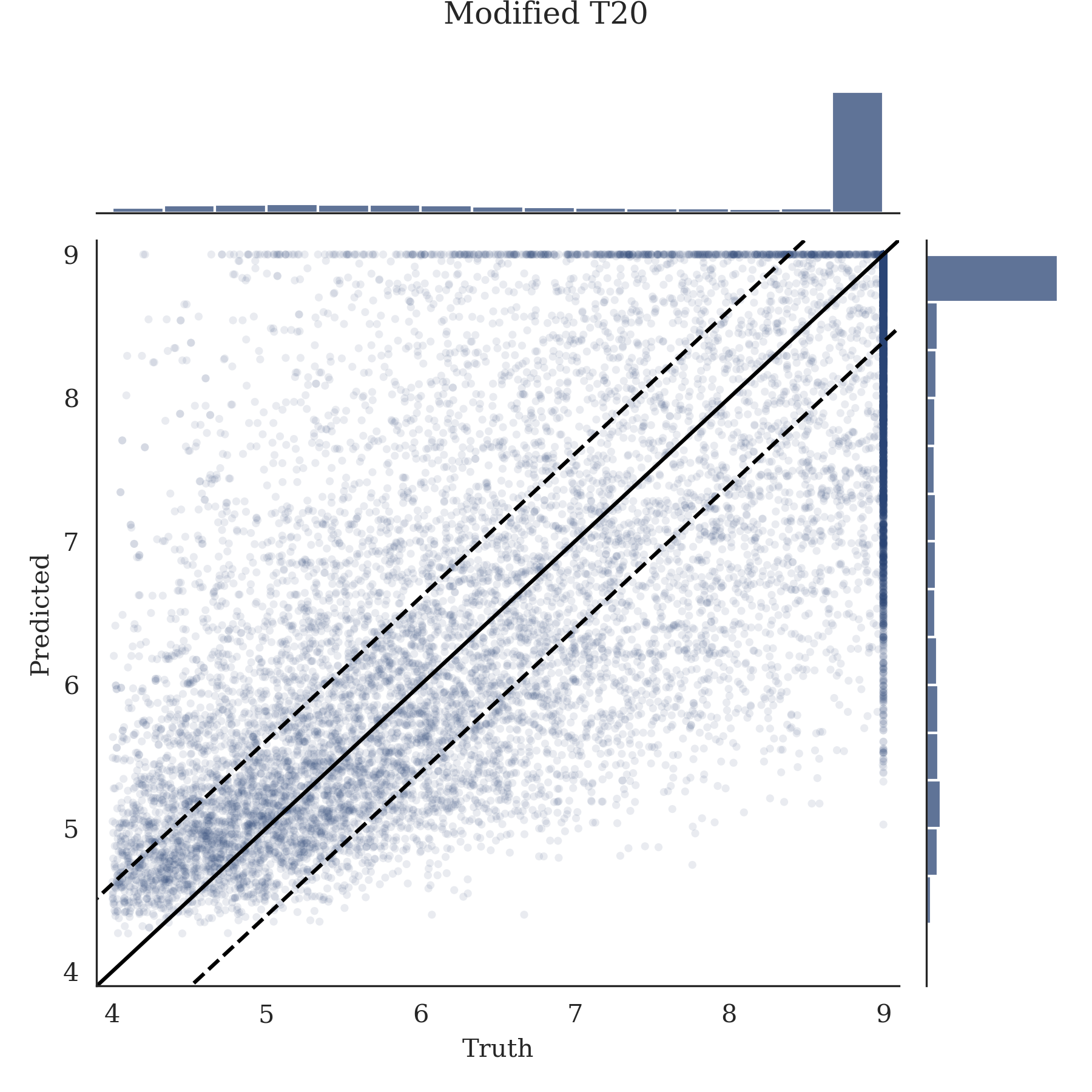}  \\
    \small The model of Obertas et al. (2017) &
      \small The model of Petit et al. (2020) &
      \small Tamayo et al. (2020), retrained for regression \\
    \includegraphics[width=\tmpwidth,trim={0 0 0 20pt},clip]{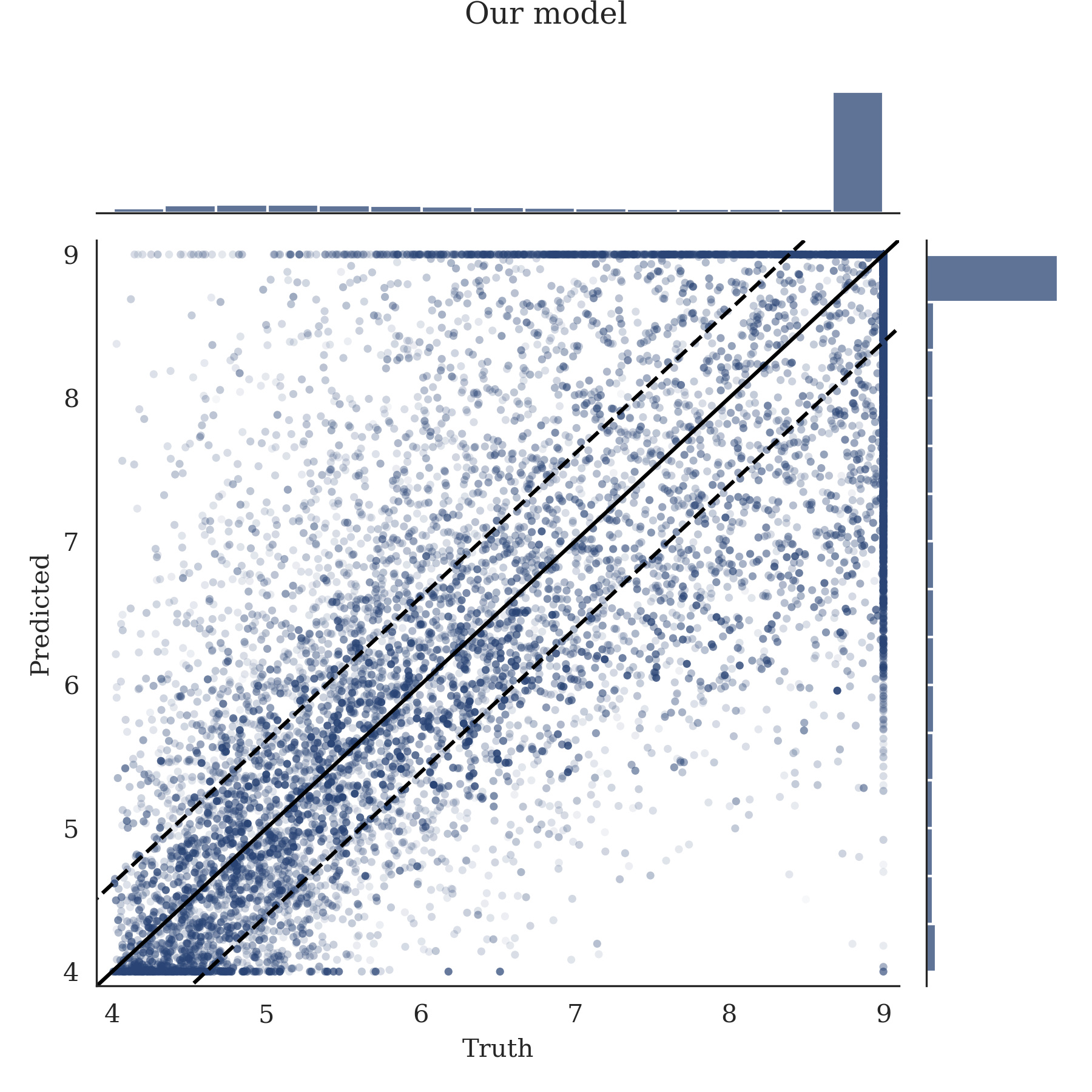} &
      \includegraphics[width=\tmpwidth,trim={0 0 0 20pt},clip]{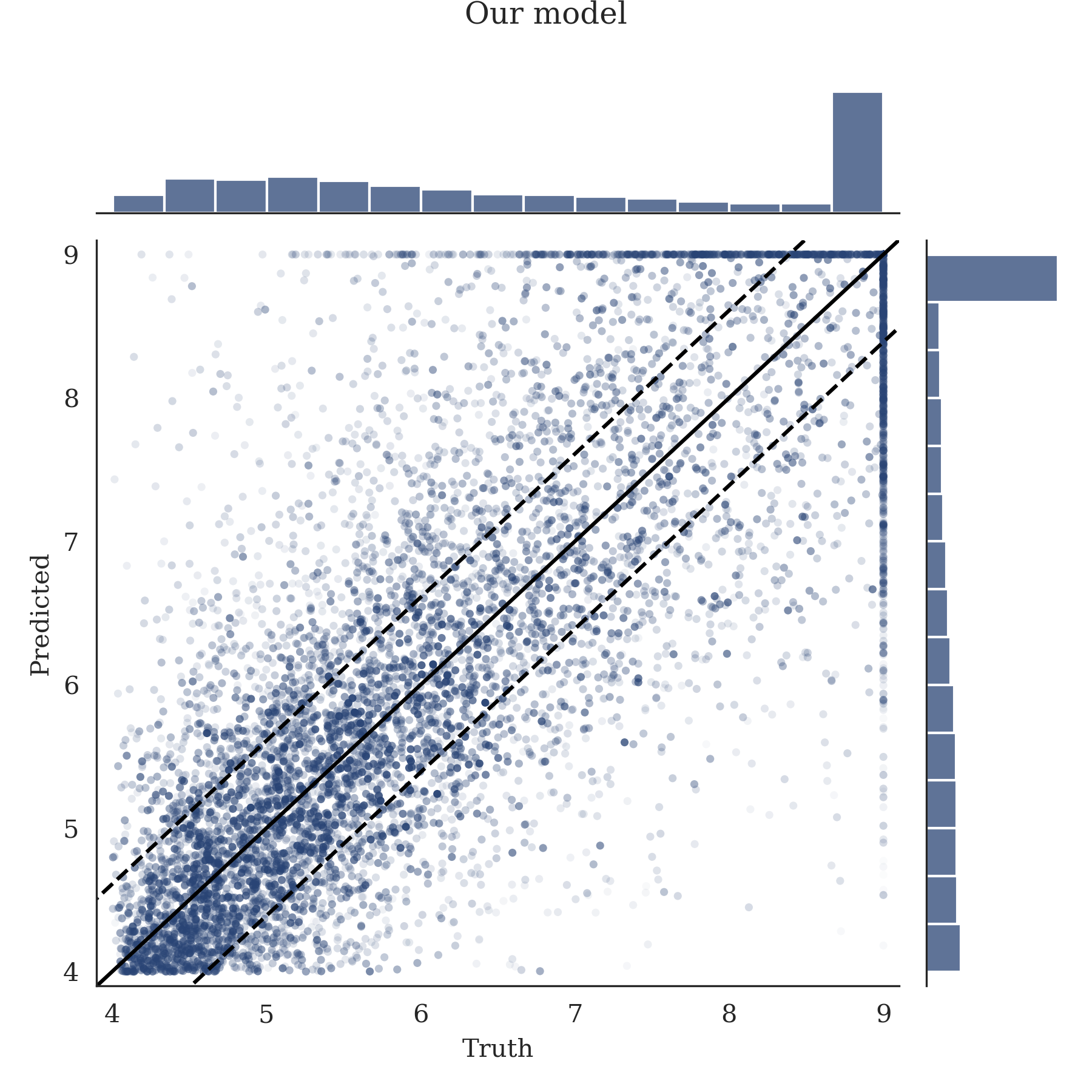} &
    \includegraphics[width=\tmpwidth,trim={0 0 0 20pt},clip]{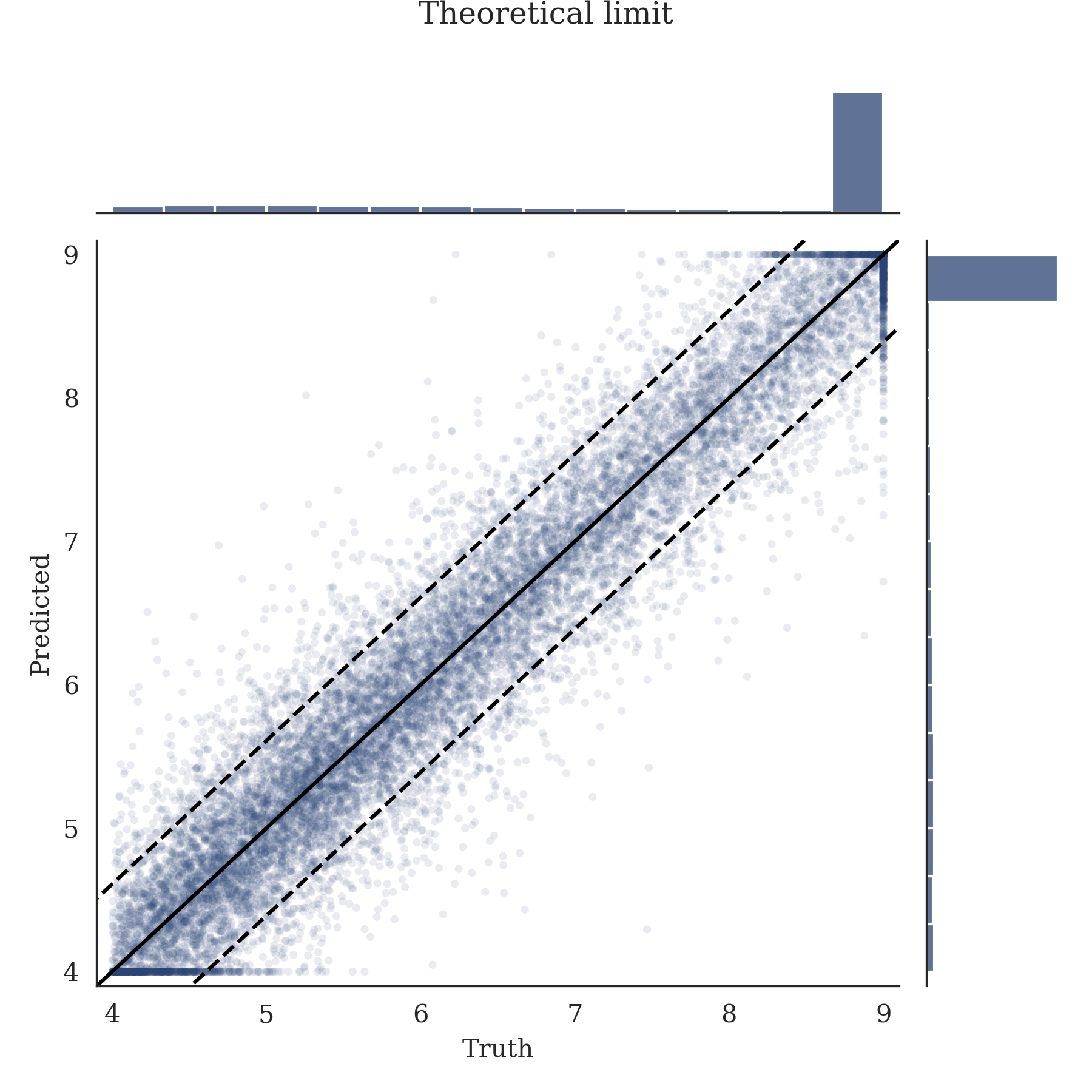} \\
      \small Our model &
    \small Our model, applied to the resonant dataset & Theoretical limit \\
  \end{tabular}
\caption{Density plots showing the predicted
    versus true instability times for various models on the
    random dataset.
    All predictions outside of $T\in[4, 9]$ are moved to the
    edge of the range.
    For the plot showing the predictions of our model,
    transparency shows relative model-predicted SNR.
    The theoretical limit, using the numerical experiments of
    \protect\cite{hussain19},
    is given in the lower-right plot. The \SI{0.43}{dex}
    RMSE average from this is used to give the dotted contours in all plots.
    }
\label{fig:reg}
\end{figure*}

First, many authors have run N-body integrations along low-dimensional cuts through the parameter space of compact orbital configurations, and fit simple functional forms to the resulting trends in instability times.
For example, several studies have highlighted the steep dependence on interplanetary separation, while fixing orbits to be coplanar and initially circular, and planets to be equal mass and equally separated from one another \citep{chambers96, marzari02, faber07, smith09, obertas17}.
We compare the performance of the fit from such a study \cite{obertas17}, using five equally spaced Earth-mass planets (mass ratio $\approx 3 \times 10^{-6}$) on our random test set in the top left panel of \cref{fig:reg}, with a resulting RMSE of 2.41.
Followup studies have incorporated the effect of finite inclinations and eccentricities \citep{yoshinaga99, zhou07, funk10, wu19}, but consider equal initial eccentricities, planetary masses, etc. in order to fit simple models. 
We conclude that while such controlled experiments yield insight into the underlying dynamics \citep{zhou07, quillen11, yalinewich19}, instability times depend sensitively on masses and several orbital parameters, rendering empirical fits to low-dimensional cuts in the parameter space of limited applicability.
In the following section we perform the converse generalization test, where we ask our model to instead predict on the N-body simulations used for the fit by \cite{obertas17}, and find good agreement.

Second, previous authors have developed analytical instability time estimates from first principles.
These have been most successful in the limit of initially circular orbits, where three-body MMRs have been identified as the dominant driver of chaos \cite{quillen11}.
Recent work \cite{petit20} has extended this theory to provide accurate instability time estimates. 
We will compare the predictions of our model to this limit of initially circular orbits in the next section.
Here we simply emphasize the point by \cite{petit20} that such predictions perform poorly at finite eccentricities (top middle panel of \cref{fig:reg}), likely due to the dominant effects of stronger two-body MMRs.
The fact that the analytic model predicts the vast majority of systems to be stable implies that most of our test configurations would be stable on circular orbits, but that finite orbital eccentricities strongly modulate instability times.

The final approach is to make predictions across the high-dimensional space of orbital configurations using machine learning \citep{tamayo16, tamayo20}.
We consider two variants of \cite{tamayo20} adapted for regression.
The first, labeled `Tamayo et al. (2020)', is to simply use an identical model as \cite{tamayo20}, but map the probability estimates of stability past $10^9$ orbits through  an inverse cumulative distribution of a log-normal with an optimized constant standard deviation.
The second, labeled
`Modified Tamayo+20', the model is an XGBoost \cite{xgboost} regression model (rather than classification) retrained on the same features as used in \cite{tamayo20}.

We find that our model achieves similar performance to the Modified Tamayo+20 variant (top right panel of \cref{fig:reg}, \cref{tbl:reg}), though the latter exhibits significant bias. 
We quantify this bias for each model in the range
$T\in(4, 5)$ and $T\in (8, 9)$.
As is evident in \cref{tbl:reg}
as well as \cref{fig:reg}, the model introduced
in this work exhibits significantly reduced bias compared to other models. 
Including SNR weighting further reduces bias.
Bias is a measure of the generalizability of a model to out-of-distribution data
\citep[see chapter 7 of][]{varbias}, so is an important metric for understanding
how these predictive models will extrapolate to new data.
Our model achieves predictions that are more
than two orders of magnitude
more accurate than the analytic models in each case:
e.g., $10^{2.41/1.09}\approx162\times$ when comparing our SNR-weighted
model with \cite{obertas17} on the random dataset.

Finally, we can make a comparison
to the original classification model of \cite{tamayo20} by using our regression
model as a form of classifier. 
We count the fraction of samples above $T=9$ as the probability a given system is stable,  
and measure the performance of the classifier with the area under
the receiver operating characteristic curve (AUC ROC) for a range of threshold probabilities for stability (\cref{tbl:reg}).

\subsection{5-planet generalization with comparison}
\label{sec:fiveplanetresults}

As a second generalization test of our model, we compare its performance on the limiting case considered by \cite{obertas17}.
This case of five equally spaced, Earth-mass planets on initially circular and coplanar orbits differs significantly from our training set of resonant and near resonant, eccentric and inclined configurations of three planets with unequal masses.
This dataset contains 17500 simulations numerically integrated for $10^{10}$ orbits \citep{obertas17}. 
This generalization to a limiting set of higher multiplicity configurations provides a stringent test of whether the model has learned features of the dynamics or whether it is naively interpolating across the distribution of orbital configurations present in our training dataset.

To extend our three-planet predictions to higher multiplicity systems, we perform the same short integration for all planets, but pass time series for each
adjacent trio of planets to the model separately.
The model samples a single instability time for each adjacent trio,
and the minimum across this set is adopted as the instability time for the
system, as an estimate of the time for the first trio to go unstable. 
This procedure is then repeated, and we record the median and confidence intervals of the resultant distribution in $T$.
Such a reduction of compact multi-planet systems to sets of adjacent trios has been proposed on theoretical \citep{quillen11, petit20} as well as empirical \citep{tamayo20} grounds. This is motivated by the fact that the perturbative effects of planets on one another fall off exponentially with separation \citep{quillen11, petit20}, so non-adjacent interactions can largely be ignored. 

\begin{figure*}[!ht]
    \centering
    \includegraphics[width=0.8\textwidth]{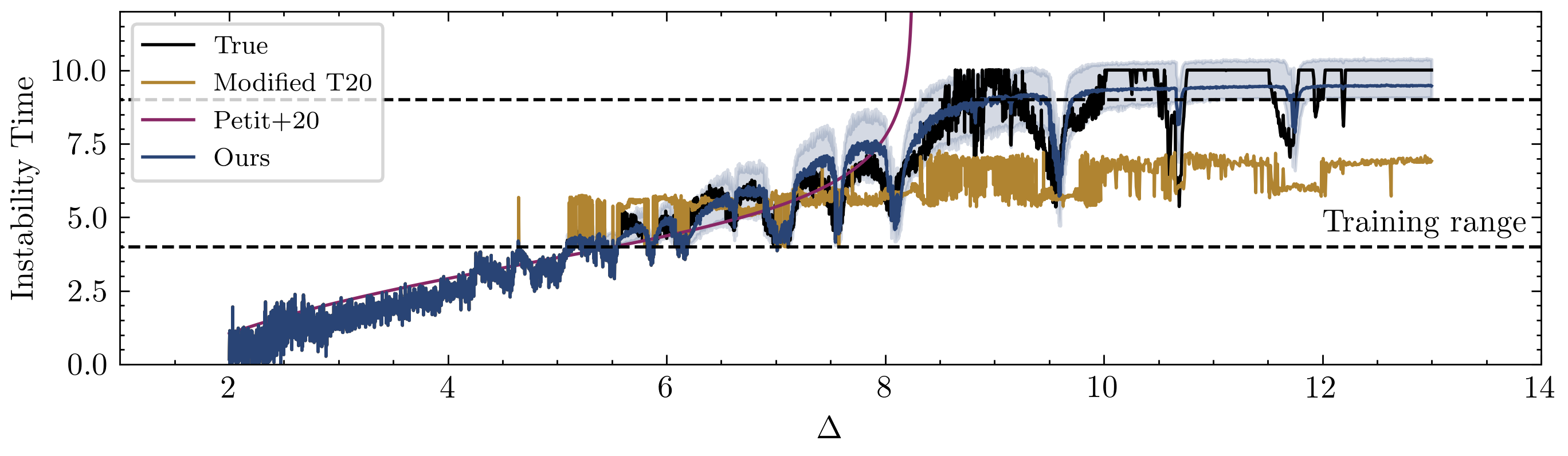}
    \caption{The median instability time
    predictions of our model for the five-planet systems used in
    \protect\cite{obertas17}.
    These systems have fixed interplanetary separation
    between each body, which is labeled on the x-axis.
    Errorbars fill out the 68\% confidence interval.
    The predictions from \protect\cite{petit20} and \cite{tamayo20} 
    are overplotted. Residuals are shown in the supplementary material.
    }%
    \label{fig:five}
\end{figure*}

The predictions can be seen in \cref{fig:five}, and are remarkably accurate despite our
model never seeing a system with five planets during training.
We overplot the analytical result of \cite{petit20} in magenta developed from first principles for such cases with initially circular orbits, including a manual correction for five-planet systems.
Our model captures the same overall trend, but additionally identifies the various dips, which correspond to locations of different MMRs \citep{obertas17}.
We emphasize that our model was trained on the general eccentric case where the magenta model of \cite{petit20} does not apply (\cref{fig:reg}), yet the generalization to this limiting case is excellent.
In addition to matching the overall trend of \cite{petit20}, our model captures the additional instability time modulations at MMRs, as can be seen more clearly in the residuals in \cref{fig:five_comp} of \materials.
Additionally, our new model generalizes much better than the predictions of the modified regression model based on \cite{tamayo20} based on manually engineered features (gold).

\subsection{Interpretation}
\label{sec:interp}

In industry machine learning, one
is focused on making predictions as accurate as possible,
even at the expense of a more interpretable model.
However, in physics, we are fundamentally
interested in understanding problems from first principles.

\begin{figure}[!ht]
    \centering
    \includegraphics[width=0.9\linewidth]{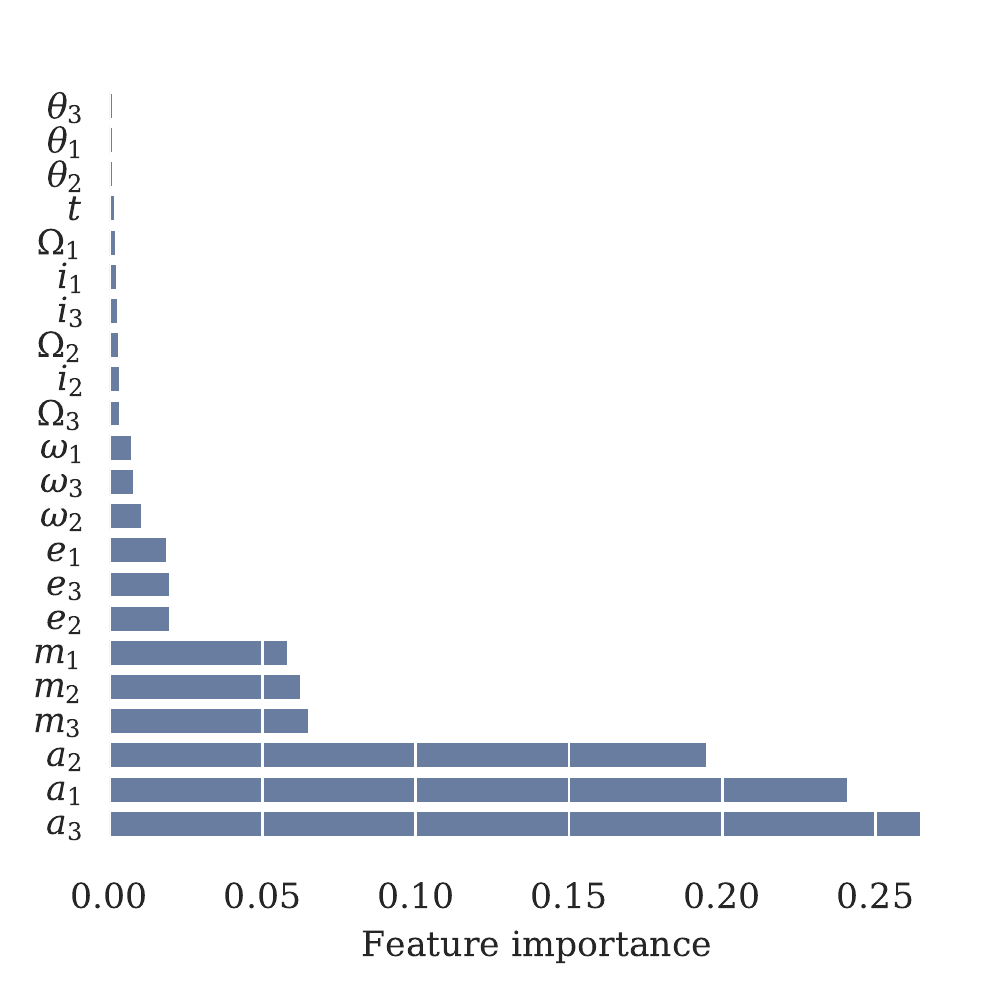}
    \caption{Feature importances in the model, calculated
    as the root-mean-square of the gradients of the model's output
    with respect to the input features, and normalized.}%
    \label{fig:importances}
\end{figure}

Obtaining such an explicit interpretation of our model
will be difficult. 
However, as a first step we consider the feature importances of our model: what orbital elements
is it using to make predictions, and does this align with expectations?
To do this feature analysis, we exploit the differentiability of
our model with respect to its input.
We calculate the gradient of the predicted $\mu$ value
our model with respect to the input instantaneous
orbital elements; this is also referred to as a ``saliency map''
\cite{deeplearn}.
This gives
us a multi-dimensional array over feature, simulation, time step, and model, representing
how much the predicted $\mu$ value will change should that feature be infinitesimally increased.
We compute the variance of the gradients
over time and each simulation, and then average
these variances over sampled network weights $\theta$.
This gives us a coarse representation of the importance of each feature.
We show this visually in \cref{fig:importances}.

To compare these importances to other work,
\cite{tamayo20} argue empirically that the short-timescale instabilities we probe here in compact systems are driven specifically by the interactions between MMRs. 
A classical result of celestial mechanics is that in the absence of such MMRs, the long-term dynamics keeps the semi-major axes fixed.
Variations in the semi-major axes during the short integrations thus act as a proxy for the importance of nearby MMRs \citep{tamayo16}, and we see that indeed, the semi-major axes exhibit the highest feature importance in our model \cref{fig:importances}.
The fact that the model ascribes comparable feature importance to any given orbital element for each of the three planets also suggests a physically-reasonable model.

We note that there is a small but nonzero
significance of the ``Time'' instantaneous feature.
This can be interpreted as being important because the model takes the first $10^4$ orbits
as input, and can predict instability for the system as low as $10^{4.1}$.
Thus, the orbital parameters given at $10^4$ orbits may be more important
than the orbital parameters at $10^0$ orbits for predicting such unstable systems,
and thus the time feature is used.
The time feature would be less important
for a system that goes unstable near $10^9$ orbits, as the relative importance of
the system's parameters at $10^4$ orbits 
is comparable to that at $10^0$ orbit.

Because we chose to structure our model to take means and variances of the times series of the learned transformed variables \cref{fig:graphical}, it may be possible to extract explicit interpretations of our model via symbolic regression.
Given that our approach is structurally similar to that of a graph neural network
\cite{gnn}, the frameworks of \cite{symbolicgn1,symbolicgn2,pysr} would be particularly applicable.
This would be done by finding analytic forms for $f_1$, representing
each of the transformed variables, and then finding an analytic form
for $f_2$, to compute the instability time given the transformed variables.
This type of explicit analysis of the model will be considered in future work.

\section{Conclusion}
We have described a probabilistic machine learning model---a Bayesian neural network---that
can accurately predict a distribution over possible instability times
for a given compact multi-planet exoplanet system.
Our model is trained on the raw orbital parameters of a multi-planet
system over a short integration, and learns its own instability metrics.
This is contrasted by previous machine learning approaches which have given their models hand-designed instability metrics based on specialized domain knowledge.

Our model is more than two orders of magnitude more accurate at predicting instability times than analytical estimators, while also reducing the bias of existing learned models by nearly a factor of three.
We also demonstrate that our model generalizes robustly to five-planet configurations effectively drawn from a one-dimensional cut through the broad parameter space used to train the model.
This improves on the estimates of analytic and other learned models, despite our model only being trained on compact three-planet systems.

Our model's computational speedup over N-body integrations by a factor of up to $10^5$ enables a broad range of applications, such as using stability constraints to rule out unphysical configurations and constrain orbital parameters \citep{tamayo20b}, and to develop efficient terrestrial planet formation models.
Towards this end, our model will be made publicly available through the \texttt{SPOCK}\footnote{\url{https://github.com/dtamayo/spock}} package, with training code also
available in a separate repo\footnote{\url{https://github.com/MilesCranmer/bnn_chaos_model}}.

\paragraph{Acknowledgements}

Miles Cranmer would like to thank Andrew Gordon Wilson and Dan Foreman-Mackey
for advice on Bayesian deep learning techniques,
and Andrew Gordon Wilson, Dan Foreman-Mackey, and
Sebastian Wagner-Carena for comments on a
draft of this paper.
Philip Armitage, Shirley Ho, and David Spergel’s work is supported by the Simons Foundation.
This work made use of several Python software
packages:
\texttt{numpy} \cite{numpy},
\texttt{scipy} \cite{scipy},
\texttt{sklearn} \cite{sklearn},
\texttt{jupyter} \cite{jupyter},
\texttt{matplotlib} \cite{matplotlib},
\texttt{pandas} \cite{pandas},
\texttt{torch} \cite{torch},
\texttt{pytorch-lightning} \cite{lightning},
and \texttt{tensorflow} \cite{tensorflow}.

\bibliography{main}

\clearpage

\onecolumn

\appendix
\section*{Appendix}

\begin{figure}[!ht]
    \centering
    \begin{tabular}{cc}
    \includegraphics[width=.4\linewidth,trim={0 0 0 20pt},clip]{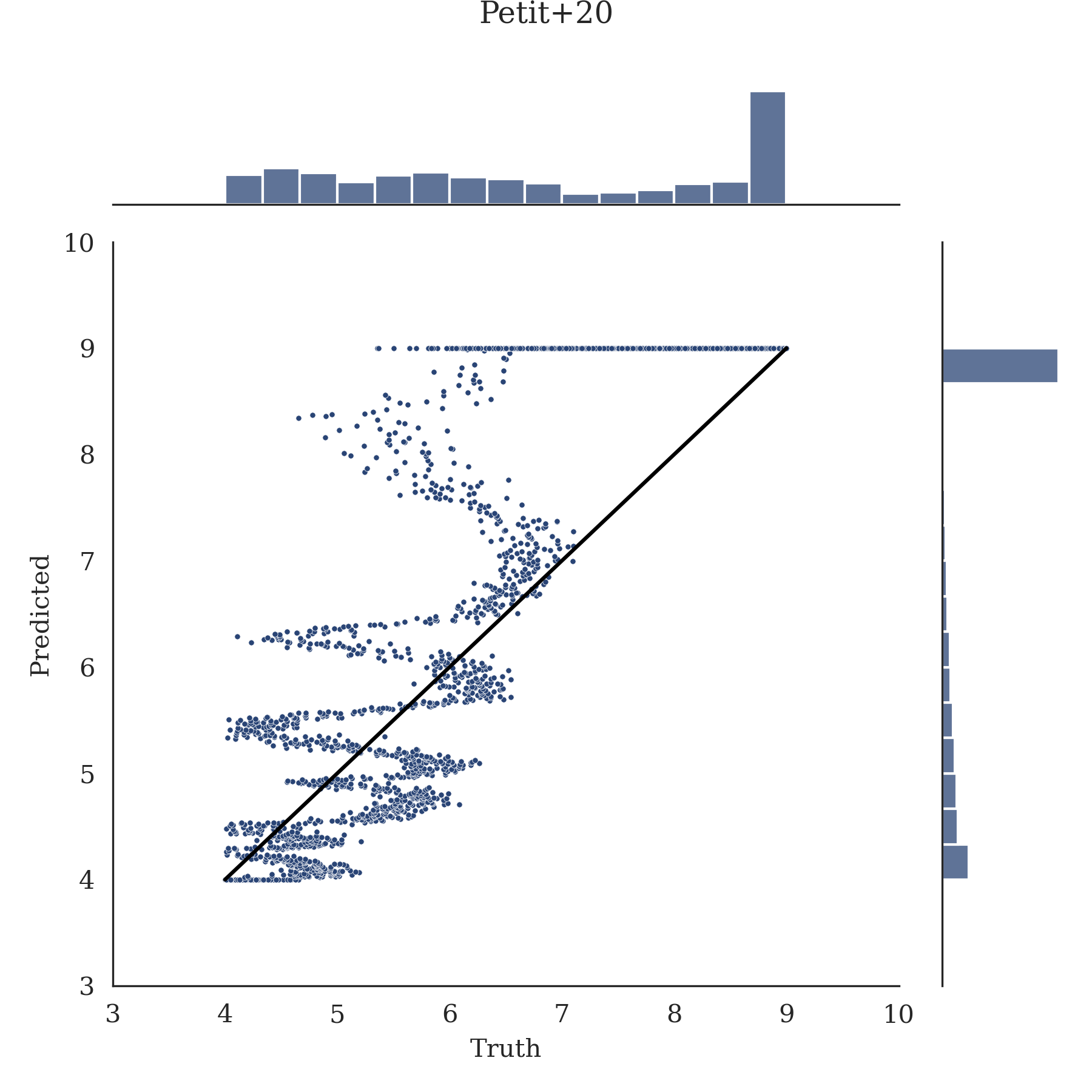} &
    \includegraphics[width=.4\linewidth,trim={0 0 0 20pt},clip]{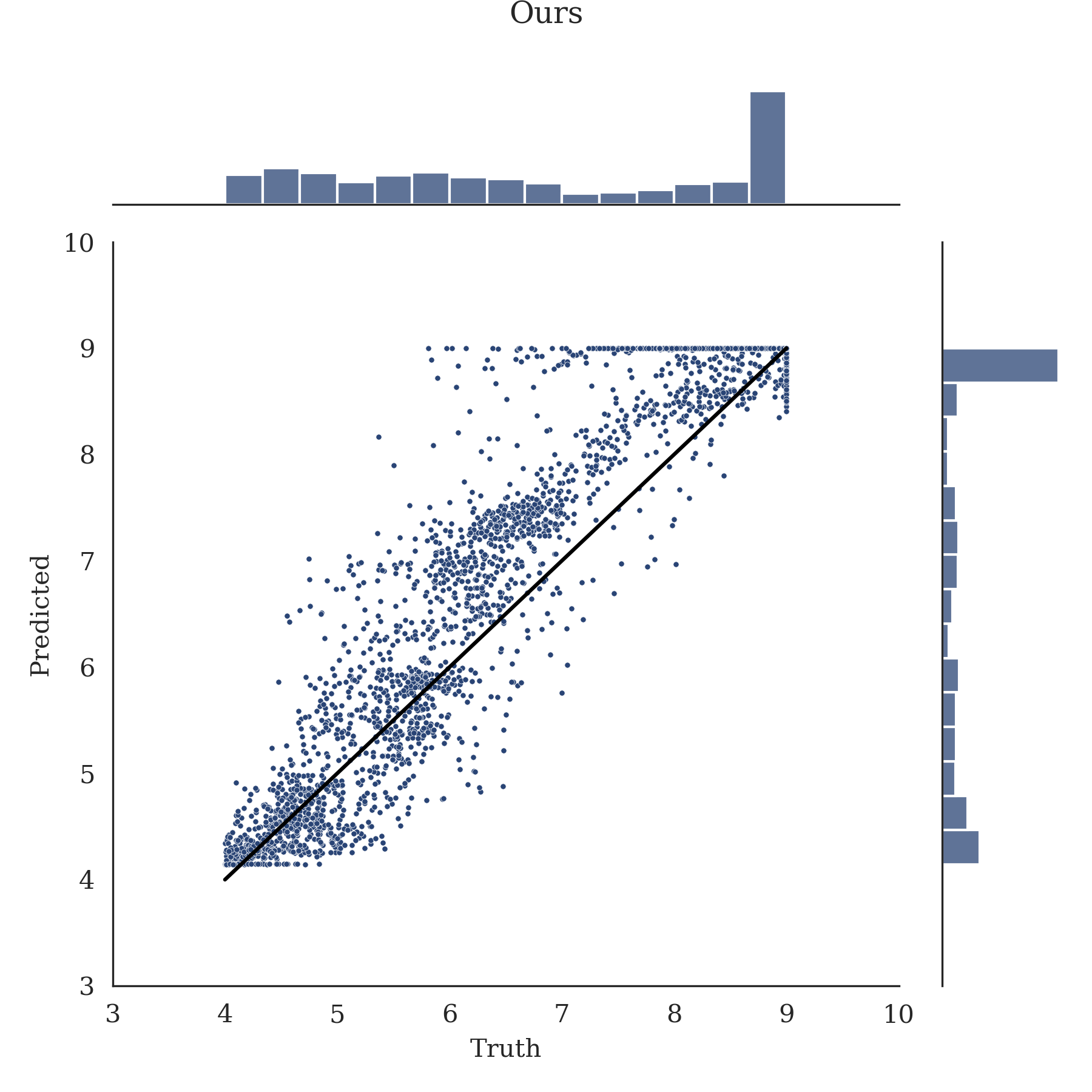} \\
    Petit et al. (2020) & Our model
    \end{tabular}
    \caption{Residuals for predictions
    of instability time on the 5-planet dataset
    in \cref{fig:five}.}
    \label{fig:five_comp}
\end{figure}

\paragraph{Parameters}
A planetary orbit can be described with six coordinates
at any given time. We choose to use
eccentricity ($e$), semi-major axis ($a$, normalized to the initial
innermost planet's value), inclination ($i$),
longitude of the ascending node ($\Omega$), longitude of pericenter ($\varpi$),
and true longitude ($\theta$).
We also pass the mass of each planet normalized to the star's mass
to each call of $f_1$ in \cref{eqn:modeleqn2}.
For each of the angular coordinates excluding inclination,
we split into two values---one
for the sine and one for the cosine of the value---before
passing to the first neural network.

\paragraph{Likelihoods}
Here we give a mathematical derivation of the likelihood
used to train our model.
Our goal is to estimate the distribution
$P(T|X)$, for a timeseries data $X \in \mathbb{R}^{3 + 19 n_t}$. %
We construct a probabilistic model defined by
\cref{eqn:modeleqn,eqn:modeleqn2,eqn:modeleqn3,eqn:modeleqn4,eqn:modeleqn5},
with parameters $\theta\in\mathbb{R}^m$, where $m$ is the total number of parameters,
which takes a timeseries for 3 planets, $X$,
and produces a normal probability distribution over $T$,
parametrized by 2 scalars: $(\mu_\theta(X), \sigma^2_\theta(X))$.
$\mu_\theta$ is the center of the instability time,
and $\sigma_\theta$ is the standard deviation in that estimate.

The distribution over $T$ parametrized by the model is equal to:
\begin{equation}\label{eqn:like}
    P(T|X, \theta) = \left\{
        \begin{array}{cc}
            \frac{A(\mu_\theta(X), \sigma^2_\theta(X))}{\sqrt{2\pi}\sigma_\theta(X)}\exp(-\frac{(T - \mu_\theta(X))^2}{2 \sigma_\theta(X)^2}),  & T< 9\\
        \\
        \frac{1}{2}\left(
        1 + \erf\left(\frac{\mu_\theta(X)-9}{\sqrt{2}\sigma_\theta(X)}\right)
        \right) P(T|T\geq 9), & T \geq 9.
        \end{array}
        \right.
\end{equation}
This distribution is motivated by two things. First, as in the paper \cite{hussain19},
exoplanet instability times usually follow a normal distribution in logarithmic
instability time, regardless of how large this time is. Therefore,
we predict a normal distribution in $T$ for times under $T<9$.
Second, due to computational costs,
we only simulate systems up to $10^9$ orbits, hence we use a model that is independent
of $P(T|T\geq 9)$.
We calculate the cumulative probability of the normal distribution
falling $T\geq9$ to calculate the probability of the value being stable.
Here, $A$ is a normalization function from the fact that we cut off the probability
at $T=4$. Thus,

\begin{equation}
    A(\mu_\theta(X), \sigma^2_\theta(X))=\frac{2}{1+\erf\left(\frac{\mu_\theta(X) - 4}{\sqrt{2 \sigma^2_\theta(X)}}\right)}.
\end{equation}
This term, from our prior that $T>4$,
helps remove bias from our model, as can be seen in \cref{fig:reg}. Without
this term in the model, we would be artificially
punishing predictions at low $T$ values.

\begin{figure}[!ht]
    \centering
    \includegraphics[width=0.45\linewidth]{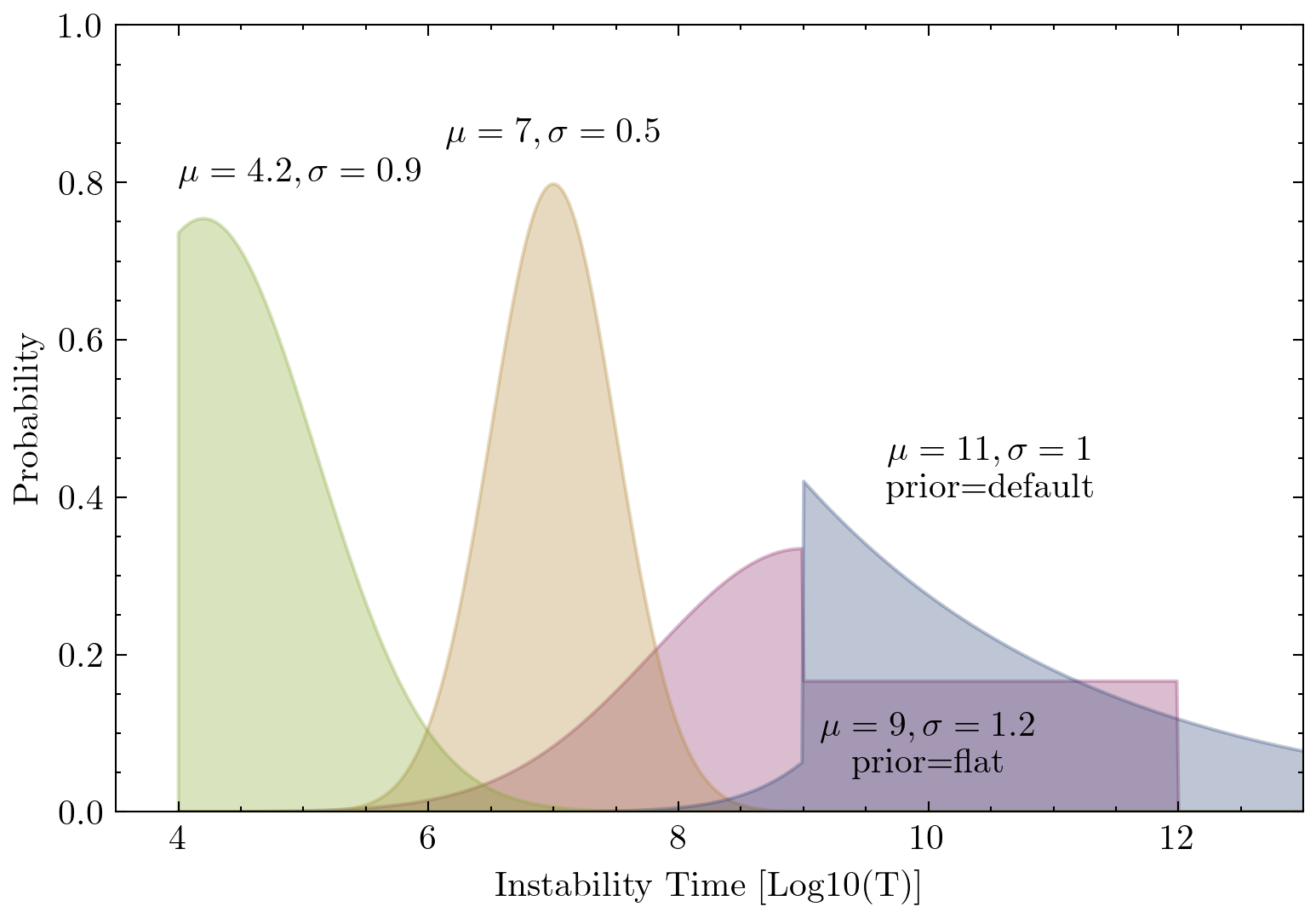}
    \caption{Example likelihoods for various choices
    of $\mu, \sigma^2$ corresponding to \cref{eqn:like}. Example priors are given for configurations
    stable past $10^9$, though note that these have no effect on the learned model.}%
    \label{fig:like}
\end{figure}

Assuming we produce a pointwise dataset $D=\{(T_i, X_i)\}_{i=1:N}$ via numerical
integration, where $T_i \equiv 9$ indicates that the system is stable beyond
$10^9$ orbits, the log-likelihood for this model is equal to:
\begin{equation}\label{eqn:sumlike}
    \log P(D|\theta) \propto \sum_i
    \left\{
        \begin{array}{cc}
            -\frac{(T_i - \mu_\theta(X_i))^2}{2 \sigma_\theta(X_i)^2} - \log(\sigma_\theta(X_i)) &\\
            - \log\left( 1+\erf\left(\frac{\mu_\theta(X_i) - 4}{\sqrt{2\sigma^2_\theta(X_i)}}\right) \right),
            & T_i< 9\\
            \log\left(1+
            \erf\left(\frac{\mu_\theta(X_i)-9}{\sqrt{2\sigma^2_\theta(X_i)}}\right)
            \right), & T_i \equiv 9,
        \end{array}
        \right.
\end{equation}
assuming a fixed prior $P(T|T\geq 9)$.
Note how this decouples the loss
for the stable values $T_i\geq 9$ from the prior $P(T|T\geq9)$,
meaning the choice of prior will have no effect on our model, and
can be chosen by a user after training.
Examples of this are plotted in \cref{fig:like}.

Now, we also marginalize the model parameters $\theta$,
to incorporate epistemic uncertainty, and account for model
biases due to the random seed used in initializing and training the model.
We proceed as follows:
\begin{align}
    P(T|X, D) &\propto \int P(T, \theta|X, D)  d\theta\\
    &\propto \int P(T|X, D, \theta) P(\theta|X, D)  d\theta\\
    &\propto \int P(T|X, \theta) P(\theta|D)  d\theta.
\end{align}
We first maximize the likelihood of the model, to find $P(\theta|D)$.
We factor the joint probability using \cref{fig:graphical}, and proceed:
\begin{figure}[!ht]
\begin{center}
    \begin{tikzpicture}
        \node[obs] (x) {$\vx_{ij}$} ; %
        \factor[right=of x] {f1} {below:$f_1$} {} {};
        \node[latent, right=of f1] (y) {$\vy_{ij}$} ; %
        \factor[right=of y, xshift=0.5cm] {f4} {below:$(\mathbb{E}[\cdot], \text{Var}[\cdot])$} {} {} ; %
        \node[latent, right=of f4] (z) {$\vz_{i}$} ; %
        \node[latent, above=of f1, yshift=1cm] (t1) {$\pmb{\theta}_1$} ; %
        \node[latent, above=of t1] (tall) {$\theta$} ; %
        \factor[above=of tall] {ft} {left:$\mathcal{N}$} {} {} ; %
        \node[latent, above=of ft, xshift=-0.5cm] (mt) {$\pmb\mu_\theta$} ; %
        \node[latent, above=of ft, xshift=0.5cm] (st) {$\Sigma_\theta$} ; %
        \factor[right=of z] {f2} {below:$f_2$} {} {};
        \node[latent, above=of f2, yshift=1cm] (t2) {$\pmb{\theta}_2$} ; %
        \node[latent, yshift=-1cm, right=of f2] (m) {$\mu_i$} ; %
        \node[latent, yshift=1cm, right=of f2] (s) {$\sigma^2_i$} ; %
        \factor[right=of m, yshift=1cm] {f3} {below:$\mathcal{N}$} {} {} ; %
        \node[obs, right=of f3] (T) {$T_i$} ; %
        \factoredge {x,t1} {f1} {y};
        \factoredge {z,t2} {f2} {m,s};
        \factoredge {m,s} {f3} {T};
        \factoredge {y} {f4} {z};
        \factoredge {mt,st} {ft} {tall};
        \edge {tall} {t1,t2} ;
        \plate {p1} {(x)(y)} {For each time step $j\in\{1..n_t\}$};
        \plate {p2} {(x)(y)(z)(m)(s)(T)(p1)} {For each simulation $i\in\{1..N\}$};
        \plate {p3} {(mt)(t1)(t2)} {For each mode of the parameter surface~~~~~~~~~};
    \end{tikzpicture}
\end{center}
    \caption{Bayesian graphical model representing our inference scheme for the instability time with Bayesian
    neural networks,
    which goes from timeseries created via short-term integration ($\{\vx_{ij}\}_{j=[1:n_t]}$) to a prediction
    of a (logarithmic) instability time ($T_i$) for each simulation $i$.
    $f_1$ and $f_2$ are neural networks parametrized
    by  $\theta\equiv (\pmb{\theta}_1, \pmb{\theta}_2)$.
    A distribution over $\theta$
    is learned according to the likelihood $P(\theta)\prod_{i} P(T_i|\{\vx_{ij}\}_j; \theta)$.
    The model is given in \cref{eqn:modeleqn,eqn:modeleqn2,eqn:modeleqn3,eqn:modeleqn4,eqn:modeleqn5}.
    Notation follows that of \protect\cite{dietz}.
    Vectors are bolded and matrices are capitalized.
    }
    \label{fig:graphical}
\end{figure}
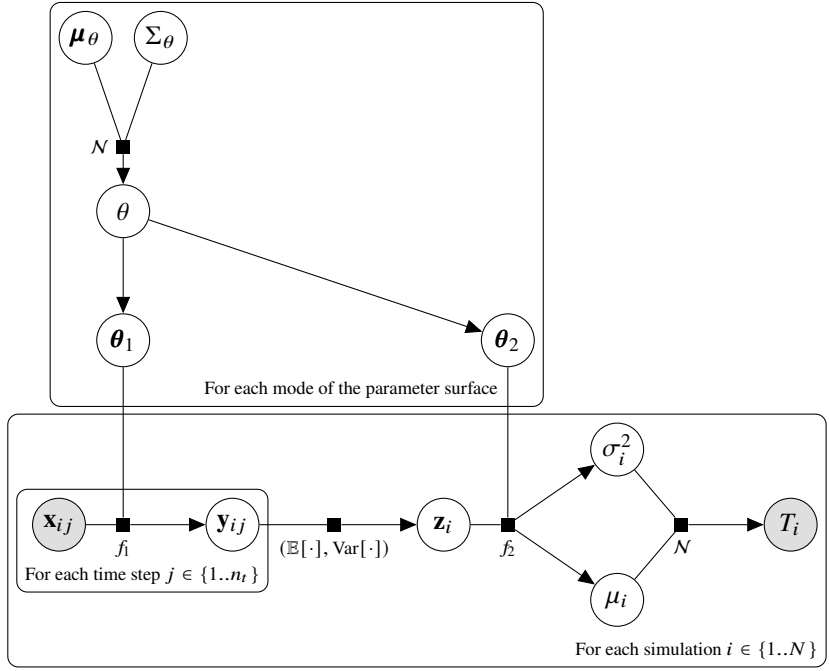
\begin{align}
    P(\theta, D) &\propto P(\theta)\prod_i \int P(X_i) P(T_i|\mu_i, \sigma^2_i)P(\mu_i, \sigma^2_i|\theta, X_i) d\mu_i d\sigma^2_i\\
    \text{Using } P(\theta | D) &\propto \frac{P(\theta, D)}{P(D)}\\
    \Rightarrow P(\theta | D) &\propto P(\theta)\prod_i \int P(T_i|\mu_i, \sigma^2_i)P(\mu_i, \sigma^2_i|\theta, X_i) d\mu_i d\sigma^2_i,
\end{align}
where $P(T_i|\mu_i, \sigma^2_i)$ is given by \cref{eqn:like}
as the log-likelihood for our model,
and $P(\mu_i, \sigma^2_i|\theta, X_i)d\mu_i d\sigma^2_i$
is our forward model given \cref{eqn:modeleqn2,eqn:modeleqn3,eqn:modeleqn4,eqn:modeleqn5}.
Finally, we can write down the loss of our model, our function to minimize,
as the negative logarithm of this:
\begin{align}
    \text{Loss}(\theta) &= -\log(P(\theta)) -
    \sum_i
    \mathbb{E}_{(\mu_i, \sigma^2_i)\sim f(X_i;\theta)} \log(P(T_i|\mu_i, \sigma^2_i)),
\end{align}
where $f(X_i;\theta)$ is the combined
model \cref{eqn:modeleqn2,eqn:modeleqn3,eqn:modeleqn4,eqn:modeleqn5}
for a given $\theta$, and $\mathbb{E}$ is used to refer to the
fact that $\vb z$ is sampled in \cref{eqn:modeleqn3}, so we average the loss
over samples.
We set $P(\theta)$ equal to a zero-centered
uninformative Gaussian prior over the parameters.
If this were a neural density estimator
instead of a full Bayesian Neural Network,
we would minimize this for a single value of $\theta$.
Alternatively, we can sample $\theta \sim P(\theta|D)$
with a Bayesian Neural Network (BNN) algorithm. We use the
MultiSWAG algorithm to do this \cite{multiswag},
as described in the main text, and find the distribution
$P_\text{MultiSWAG}(\theta)\approx P(\theta|D)$.

\paragraph{Training}
We have $n_t=100$ uniformly spaced samples of our integration
over $10,000$ initial orbits (the unit orbit is
the initial period of the innermost planet).
During training, we randomly select between 5 and 100 time steps,
with replacement, to feed to the model.
This is a type of data augmentation
that improves generalization of the model.

Since we are working with a varying number of orbit samples,
we also sample the averages and variances from Gaussians over their
frequentist distributions: $\mu\pm \frac{\text{Var}_t[\mathbf{y}_t]}{n_t}$ for the mean,
and $\sigma^2\pm \frac{2 \text{Var}_t[\mathbf{y}_t]^2}{(n_t - 1)}$ for the variance,
where $n_t$ is the number of samples, and $\text{Var}_t$ is the sample variance.
This will naturally allow the model to grow increasingly certain
if a longer time series is given as input, since the averages and variances
of the transformed coordinates are less subject to small-sample uncertainty.

\paragraph{Hyperparameters}
For our final model, we set both $f_1$ and $f_2$ to be a MLPs
with ReLU activations: 1 hidden layer and $40$ hidden nodes each (i.e.,
a matrix multiplication, a ReLU, a matrix multiplication, a ReLU,
and another matrix multiplication).
The number of calculated transformed features from $f_1$
is $20$, meaning $f_2$ takes $40$ features as input.
We take 500,000 stochastic gradient descent optimization steps with random batches
of simulations with a
batch size of 2000 with a cosine-annealed learning rate \cite{onecyclelr1,onecyclelr2}
from $5\times10^{-4}$
down to $5\times10^{-8}$ 
This is followed by 50,000 additional steps at a fixed learning rate
(presumably within a mode of the posterior)
of $2.5\times 10^{-4}$, to record points of a Gaussian mode on the weight posterior.
Gradient clipping of the L2-norm
is used with a clipping value of $0.1$. A small amount
of noise is added to the input
features and summary statistics to encourage the model
to only use features that are useful: a KL-divergence loss
term is added to the loss function on this noise, with a
multiplier of $10^{-5}$ on the input and $10^{-3}$ on the summary.
This noise is not added during evaluation, only training.
We choose $5$ as the minimum number of time steps to pass to the model.
We rescale the data to have a zero mean and unit variance
for each feature (i.e., we normalize
with a mean and variance calculated over the entire training set and time series).
All of these parameters were found with the
hyperopt\footnote{\url{https://github.com/hyperopt/hyperopt}}
package with a 20\% validation set, with a smaller number of training steps and
accelerated learning rate schedule.
Finally, we train an ensemble of 30 independent models:
this represents 30 modes of the weight posterior.
For a summary of these details, see our training code at
\url{https://github.com/MilesCranmer/bnn_chaos_model}.

\begin{figure}[!ht]
    \centering
    \includegraphics[width=0.4\linewidth]{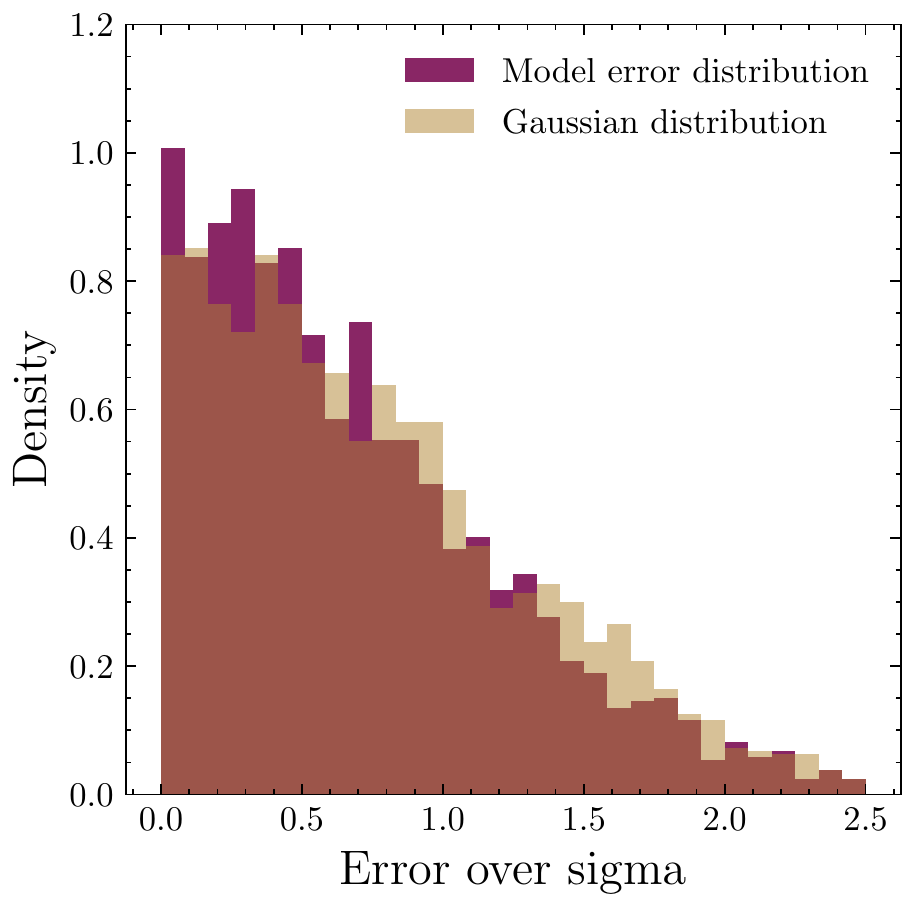}
    \caption{A comparison of the residual between our model and the
    true value with a true Gaussian of the same standard deviation,
    with the same number of samples.
    What this shows is that if our model predicts an uncertainty of $\sigma^2=1$,
    then 68\% of the true values will be within that $\sigma$ range, as expected
    for a true Gaussian distribution.
    }%
    \label{fig:exampledist}
\end{figure}

\paragraph{Approximating the cumulative distribution of a Gaussian}
Due to precision issues of 32-bit floating point numbers,
our auto-differentiation software, \texttt{torch} \cite{torch}, is increasingly
incapable of accurately calculating $\log(1+\erf(x))$ and its gradients as
$x$ decreases below $-1$. Because we heavily rely on this function in our log-likelihood for training
our model, and need to pass gradients through it, we needed to
approximate it for large negative $x$ values. Otherwise, we found
that the gradients in our model would often approach very large values
and training would fail.
We approximate this function with an analytic continuation
via symbolic regression using \texttt{PySR} \cite{pysr}.
We generate values of this function in high precision code,
and then fit analytical forms with \texttt{PySR}.
We find that the function is very
accurately approximated over $x\in[-5, -1]$ by:
\begin{align*}
\log(1+\erf(x)) \approx&  - 0.643250926022749 + 0.485660082730562*x - 0.955350621183745*x^2\\
&+ 0.00200084619923262*x^3 + 0.643278438654541*\exp(x),
\end{align*}
and this function has equivalent asymptotic properties.
We therefore use this formula in place of $\log(1+\erf(x))$ in our learning
code. \texttt{torch} code is given as follows, to replace
any appearance of $\log(1+\erf(x))$ in code:

\begin{python}
def safe_log_erf(x):
    """Compute log(1+erf(x)) with an approximate analytic continuation below -1"""
    base_mask = x < -1
    value_giving_zero = torch.zeros_like(x, device=x.device)
    x_under = torch.where(base_mask, x, value_giving_zero)
    x_over = torch.where(~base_mask, x, value_giving_zero)

    f_under = lambda x: (
         0.485660082730562*x + 0.643278438654541*torch.exp(x) +
         0.00200084619923262*x**3 - 0.643250926022749 - 0.955350621183745*x**2
    )
    f_over = lambda x: torch.log(1.0+torch.erf(x))

    return f_under(x_under) + f_over(x_over)
\end{python}

\paragraph{Theoretical limit}
In \cite{hussain19}, the authors measure
the distribution of instability times for various orbital configurations.
Taking an initial orbital configuration, the authors perturb the system by machine precision,
and measure the instability time, and repeat.
For each system, the authors then measure the mean instability time, $\mu$ (in log-space),
as well as the standard deviation, $\sigma$ (in log-space, modeled as a log-normal).
What this means is that we can
define a ``theoretical limit'' to the accuracy
with which we can predict the instability time,
and this accuracy is bounded by $\sigma$, for we cannot predict the instability time
for a given system better than within one $\sigma$ standard deviation on average.

For the purposes of this paper, we simulate an optimal estimator by taking
a particular instability time, and then making a random prediction for its instability time
within one $\sigma$ of the actual instability time.
\cite{hussain19} found that $\sigma$, while it is different for different configurations,
does not correlate to $\mu$, so for the numerical value of $\sigma$, we
simply randomly select numerical $\sigma$ values
from those released for \cite{hussain19}.
On average, these standard
deviations are \SI{0.43}{dex}.

\end{document}